%% file: main.tex
\documentclass[10pt, conference, compsocconf]{IEEEtran}
\IEEEoverridecommandlockouts
\usepackage{cite}
\usepackage{amsmath,amssymb,amsfonts}
\usepackage{algorithmic}
\usepackage{graphicx}
\usepackage{textcomp}
\usepackage{xcolor}
\def\BibTeX{{\rm B\kern-.05em{\sc i\kern-.025em b}\kern-.08em
    T\kern-.1667em\lower.7ex\hbox{E}\kern-.125emX}}

\usepackage{booktabs} 
\usepackage{makecell}
\usepackage[ruled]{algorithm2e}
\usepackage{threeparttable}
\usepackage{amssymb}

\usepackage{graphicx}
\usepackage{epstopdf}
\usepackage{multirow}
\usepackage{subfigure}
\usepackage{amsmath}
\usepackage{xspace}
\usepackage{xcolor}
\usepackage{array}
\usepackage{balance}
\PassOptionsToPackage{hyphens}{url}
\usepackage{hyperref}
\usepackage{url}
\usepackage{comment}
\usepackage{alltt}
\usepackage{array}
\graphicspath{{./figs/}}

\makeatletter

\newcommand{\squishlist}{
	\begin{list}{$\bullet$}
		{ \setlength{\itemsep}{0pt}      \setlength{\parsep}{3pt}
			\setlength{\topsep}{3pt}       \setlength{\partopsep}{0pt}
			\setlength{\leftmargin}{3.5mm} \setlength{\labelwidth}{1em}
			\setlength{\labelsep}{0.5em} } }
\newcommand{\squishend}{
\end{list}}

\begin{document}
\title{SMS Goes Nuclear: Fortifying SMS-Based MFA in Online Account Ecosystem}

\author{\IEEEauthorblockN{Weizhao Jin\IEEEauthorrefmark{1}\textsuperscript{\textsection},
Xiaoyu Ji\IEEEauthorrefmark{1}\textsuperscript{\textsection}, Ruiwen He\IEEEauthorrefmark{1},
Zhou Zhuang\IEEEauthorrefmark{1}, Wenyuan Xu\IEEEauthorrefmark{1} and Yuan Tian\IEEEauthorrefmark{2}}

\IEEEauthorblockA{\IEEEauthorrefmark{1}Zhejiang University \IEEEauthorrefmark{2}University of Virginia\\
Email: weizhaoj@usc.edu, \{xji,rwhe97,zhuangzhou,wyxu\}@zju.edu.cn, yuant@virginia.edu}
}


\maketitle

\begingroup\renewcommand\thefootnote{\textsection}
\footnotetext{Co-first authors}
\endgroup

\thispagestyle{plain}
\pagestyle{plain}

\input{Abstract}

\input{Introduction}

\input{Threatmodel}
\input{Design}
\input{Measurement.tex}
\input{CaseStudy}

\input{RelatedWork}
\input{Discussion}

\input{Conclusion}

\bibliographystyle{IEEEtran}
\bibliography{mybib}

\input{appendix}

\end{document}

%% file: Abstract.tex
\begin{abstract}
	
With the rapid growth of online services, the number of online accounts proliferates. The security of a single user account no longer depends merely on its own service provider but also the accounts on other service platforms (We refer to this online account environment as Online Account Ecosystem). In this paper, we first uncover the vulnerability of Online Account Ecosystem, which stems from the defective multi-factor authentication (MFA), specifically the ones with SMS-based verification, and dependencies among accounts on different platforms. We propose Chain Reaction Attack that exploits the weakest point in Online Account Ecosystem and can ultimately compromise the most secure platform. Furthermore, we design and implement ActFort, a systematic approach to detect the vulnerability of Online Account Ecosystem by analyzing the authentication credential factors and sensitive personal information as well as evaluating the dependency relationships among online accounts. We evaluate our system on hundreds of representative online services listed in Alexa in diversified fields. Based on the analysis from ActFort, we provide several pragmatic insights into the current Online Account Ecosystem and propose several feasible countermeasures including the online account exposed information protection mechanism and the built-in authentication to fortify the security of Online Account Ecosystem.

	

	
\end{abstract}

%% file: Introduction.tex

\section{Introduction}
\label{sec:introduction}

With the rich information and functionalities associated with online accounts, the security of online accounts is increasingly critical. Recently, there are several appalling user privacy scandals and severe data breach incidents~\cite{equi,Databreach} associated with online accounts covered by the media. The Multi-factor Authentication (MFA)~\cite{owen2008multi} proposed for strengthening an individual account seems like a viable solution. 
However, this solution is still far from being a solid protection due to its limitation to a single account of a single service. The common wisdom has yet to realize the fact that different online accounts of one single user are highly coupled with each other. This could potentially form an fragile Online Account Ecosystem. The bottleneck of the security of the ecosystem resides at the weakest nodes within it, which can be characterized as lacking secure authentication steps and exposing much sensitive personal information. In practice, there is one factor that exposes this vulnerability the most, namely, the notorious insecure SMS-based authentication ~\cite{mulliner2013sms}. 

\begin{figure}[!t]
	\centering
	\includegraphics[width=0.45\textwidth]{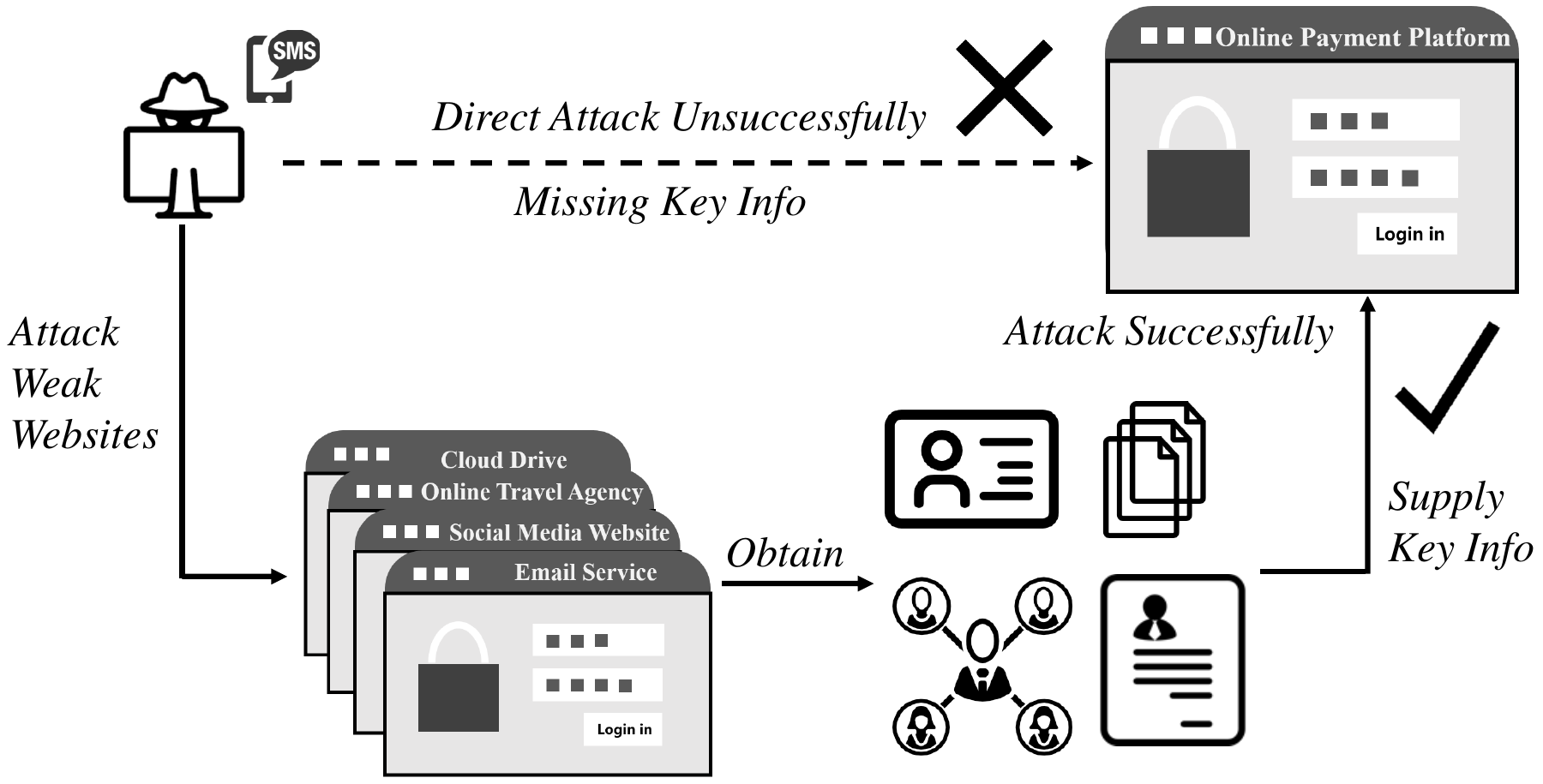}
	\setlength{\abovecaptionskip}{-0.8pt}
	\caption{Potential Online Account Ecosystem Vulnerability} 
	\label{Fig:first}
	\vspace{-0.5cm}
\end{figure} 

Although SMS-based Authentication Code (SMS Code) simplifies the authentication procedure and makes users free from forgetting the passwords, it may lead to a fatal threat to account security and user privacy that has been studied by many researchers. Several attacks have been conducted to steal SMS Code from the cellphones directly by cellphone trojans~{\cite{apvrille2010zeus}} and phishing link~{\cite{gelernter2017password}}. Moreover, some researches show that the communication between smartphones and base stations can be eavesdropped and decrypted using protocol weaknesses~{\cite{barkan2005conditional,biryukov2000real,golde2012weaponizing}}. On the basis of the previous researches, in our Chain Reaction Attack, we implemented the passive SMS Code sniffing attack and the more covert active SMS Code MitM attack to show the vulnerability of the SMS-based authentication. Fig.~\ref{Fig:first} illustrates the scenario that attackers may fail to intrude a highly secure online payment platform directly but will succeed indirectly by hacking other less secure online accounts and collecting personal information from them to gain access to the online payment platform eventually.

In this paper, we first leverage the main vulnerabilities of Online Account Ecosystem and demonstrate Chain Reaction Attacks on several real online payment platforms. Then, we design and implement ActFort, a systematic framework to formulate and analyze the vulnerabilities of the Online Account Ecosystem regrading account dependency. By analyzing hundreds of mainstream websites, we summarize some important insights existing in the current Online Account Ecosystem.

We summarized our contributions as follows:
\squishlist
\item We uncovered the essential security problem of Online Account Ecosystem which stems from the insecure SMS-based Authentication and interconnected dependencies of the online accounts.  
\item We provided a novel perspective of online account security that the security of one single service depends on the ecosystem and its "neighbours". This perspective can be extended to other attack surfaces beyond SMS Code.
\item We proposed and evaluated the ActFort on hundreds of representative online services and demonstrated the Chain Reaction Attack on real online accounts. 
\item We provided some insights into the whole ecosystem based on the measurement results.
\squishend

\noindent\textbf{Ethics.} Throughout this study, we have ensured that all of our experiments meet community ethical standards. First, We performed the attack experiments carefully and did not affect any other users. We only used our own cellphones and Internet accounts as targets. 
Second, we did not collect or obtain the personal information of any users; our small set of users only provided us with their cellphone number. Our attacks did not interact with their accounts in any ways that would reveal any additional information about them.


%% file: Threatmodel.tex
\section{Threat Overview}
\label{threat}
Most of the online account services adopt SMS Code as one of the credential factors in its login or password reset step for two reasons: accessibility~\cite{GuardianMobile,Wesoc} and convenience~\cite{zviran1990comparison,yan2004password}. In practice, the SMS-based authentication can be mainly divided into 2 types, only SMS Code or SMS Code with other credential factors (e.g. legal name and citizen ID). Unfortunately, according to our research, both authentication methods are unsafe under certain circumstances. The fundamental cause of the loophole is the reciprocal transformation of sensitive personal information and authentication credential factors among various accounts. 

It is worth noting that although we use SMS Code as our initial attack surface here, our attack can easily be extended to other factors as the major vulnerable point such as email authentication codes. The key idea of our work is to investigate account vulnerability introduced by dependencies among accounts on different platforms. 

\noindent\textbf{Chain Reaction Attack.} As described in the previous section, online accounts are still vulnerable to attacks even with the secure setup of multiple factor authentication. Chain Reaction Attack is an attack strategy that exploits the dependency among accounts from different online services and leverage the credentials and account personal information to conduct a series of attacks. Similar to the chain reactions in chemistry, our attack achieves the ultimate goal of attacking a highly-secure account by compromising less-secure service accounts first, collecting information in those accounts, and utilizing the information gathered as credentials to take control of sequential accounts. The real-world Chain Reaction Attacks will be demonstrated in Chapter V.

The goal of the attacker is to gain the control of victims' accounts. We categorized our attacks into two:
\squishlist
\item {\bfseries{Random Attack.}} The attacker aims to attack arbitrary victims nearby and has no knowledge about the victims in advance. In practice, the attack can be conducted in the airports or the railway stations which have a large flow of people and are easy for the attacker to hide and escape.
\item {\bfseries{Targeted Attack.}} The attacker aims to attack the target victim and has some knowledge about the victim in advance (e.g. the home address and the cellphone number).
\squishend 
In our threat model, the attack has no access to the internal software/hardware of the victim's cellphone or computer. In addition, we assume the attacker can stay in the same cell in the communication network. The scope of the attack is limited to scenarios where the attacker is near the victim (within a distance of hundreds of meters) because the attacker needs to either intercept the SMS Code in the same cell or hijack the victim's cellphone to complete a MitM attack. Although the attacker can remotely conduct phishing attacks to lure victims to give away SMS Codes using social engineering methods like phishing emails, this type of attack is less stealthy and requires victims' response.

%% file: Design.tex
\section{System Design}

\subsection{System Overview}

In this part, we will illustrate the design and implementation of ActFort in details. To better analyze dependency vulnerabilities in Online Account Ecosystem and explore feasible Chain Reaction Attacks, we have designed and implemented ActFort, a systematic approach to examine the potential risks of online accounts as well as generate the transformation dependency graphs of current online services. The flowchart of ActFort is mainly simplified into the Authentication Process, Personal Information Collection, Dependency Graph Generation and Strategy Output.
\begin{figure}[!t]
	\centering
	\includegraphics[width=0.45\textwidth]{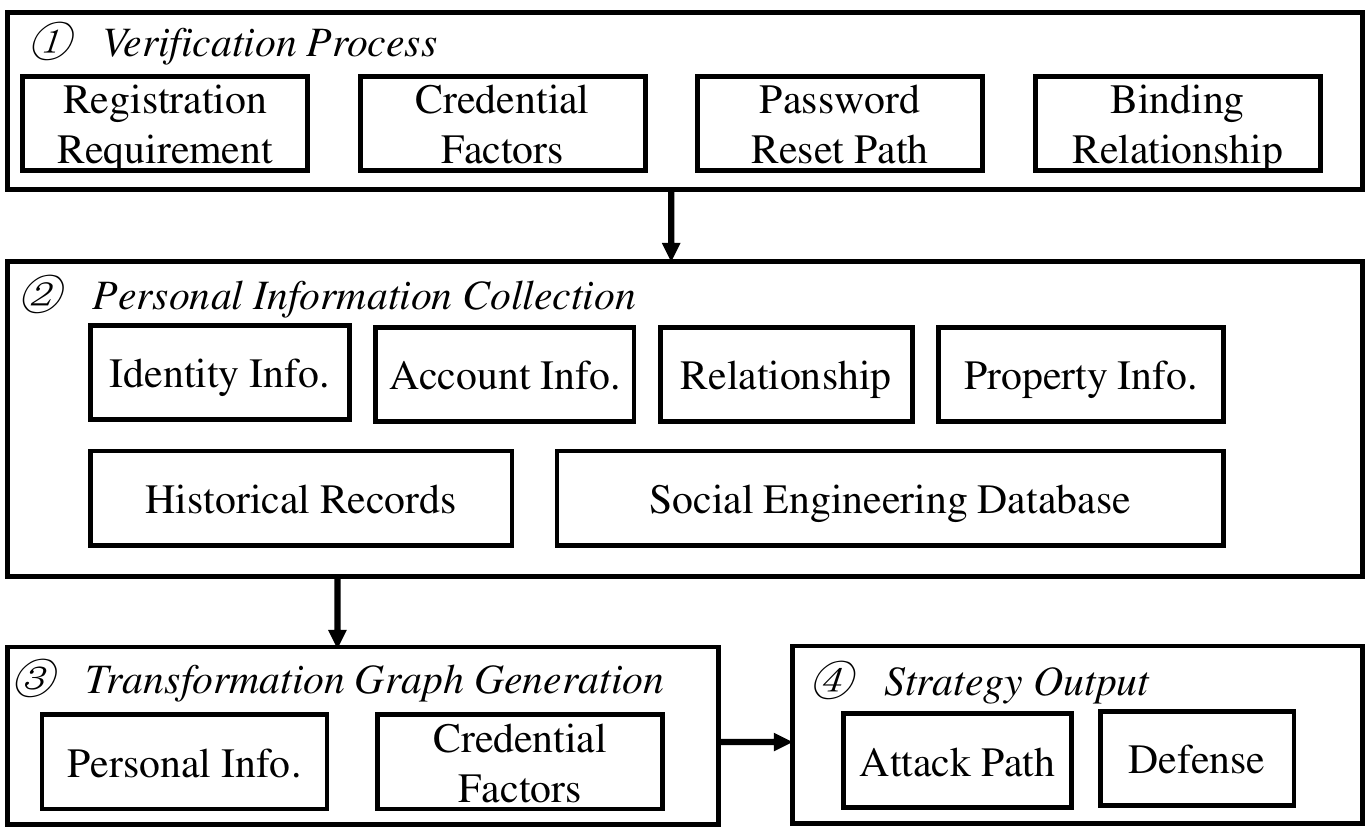}
	\setlength{\abovecaptionskip}{-1pt}
	\caption{Flowchart of ActFort.} 
	\label{Fig:ACCscanframework}
	\vspace{-0.5cm}
\end{figure}

The Authentication Process is designed for examining the registration requirements and conditions, analyzing the needed credential factors, excavating the vulnerable password reset path and determining the binding relationship for authentication.

The Personal Information Collection is utilized for collecting personal information and transferring it into structured data, which are the prerequisites for Dependency Graph Generation.

The Dependency Graph Generation is used for developing the links between credential factors and personal information originated from different online accounts.

The Strategy Output is applied for targeting the the potential vulnerabilities of online accounts and generating the potential attack path with data association rules.   

During the measurement, we input the required reset credentials for accounts and the personal information that can be obtained once the accounts are compromised to the system, then the Dependency Graph Generation will construct the links for the current Online Account Ecosystem. Once the entire dataset we have has been processed by the system, we could have a thorough analysis on the ecosystem. Additionally, if we want to add an account to the system, Strategy Output will scrutinize the existing dependency graph to find potential attack path. The design details will be discussed in each module.

\subsection{Authentication Process}
In many online account services, users can choose username, e-mail address, cellphone number or linked account to sign up with a password. Thus,  the following login or password reset can be varied due to the difference when signing up. For each online account, Authentication Process will first collect and analyze the registration requirement, recording each available signup approach and then sign up for accounts in these approaches respectively. Then Authentication Process will collect and trace the credential factors to construct the Authentication flow in each signup approach recursively.

The authentication flow construction will accept the top-down approach. The source of the authentication flow can rather be a simple login or a target action like making a payment or transfer. From the source, the flow will lead to the required steps to achieve the previous layer. To complete those steps, we might need to go further down to even lower layers until the requirement are satisfied. Note that the same credential factor could be used in different password reset paths.

In addition, the collected credential factors will be structured in order to form the paths of password reset and later be used in the transformation dependency graph generation. The binding relationship among online accounts will also be examined. 

\subsection{Personal Information Collection}

In this module, personal information in different online accounts will be collected and classified to different categories. In practice, it can be mainly divided into five categories: identity information (e.g. real names and SSN/Citizen ID), account information (e.g. email address and binding account), social relationship (e.g. friend names and family members), property information (e.g. bankcard number and savings account number), and historical records (e.g. online shopping list and chat history).  

Personal information from a compromised account becomes available credentials, building relations between accounts. This process can be considered as a transformation from personal information to credentials.

\subsection{Transformation Dependency Graph Generation}

First, we define Transformation Dependency Graph (TDG), a data structure used to represent all transformation dependency relations in the Online Account Ecosystem. We restrict our attention to transformation dependency relations in 2 categories. The first category is the relation between the credential factors and the personal information of different accounts. The other one is the linked/binding relation among the online accounts. for example, once the Gmail account is logged in, the Expedia account linked to that Gmail account can also be logged in without additional authentication.


For each node in the dependency graph, it represents an online account which contains personal information attribute ($\mathit{PIA}$) and credential factor attribute ($\mathit{CFA}$). The personal information attribute of each online account contains information relevant to \emph{real name, citizen ID, cellphone  number, e-mail address, bankcard number, address, user ID, binding account, acquaintance name, device type, and other potential authentication required information}. The credential factor attribute contains the credential factors of different authentication path for one online account. In the transformation dependency graph, there's an attacker profile ($\mathit{AP}$) which contains information about an assumed attacker's capabilities, such as SMS Code interception, social engineering database, and etc. 


We denote the credential factor attributes as inputs and the personal information attributes as outputs. For each node, the number of input depends on the available authentication paths and the number of the output rely on the personal information. Once the input and output values are identified, we construct a directed dependency graph. An edge from an output of node $\mathit{PI_{jn}}$ to an input of node $\mathit{CF_{im}}$ ($\mathit{PI_{jn}} \xrightarrow{} \mathit{CF_{im}}$) is added if one of outputs in $\mathit{PI_{jn}} $ overlap with the input value of $\mathit{CF_{im}}$ from another account: Add $\mathit{e(v_{im}, v_{jm})}$ in $\mathit{G=\{V, E\}}$, if $\mathit{{PI_{jn}} = CF_{im}}$.



%

On the basis of Transformation Dependency Graph, we can further explore the degree of the correlation of online accounts in the Transformation Dependency Graph. We first give three definitions on the correlation of online accounts.

{\bfseries{Definition 1.}} If $u$ can provide all the credential factors needed for at least one authentication path of $v$ with the attack profile, $u$ is then called the full capacity parent node of the node $v$. The directed edge from $u$ to $v$ is called strong-directivity edge. 

{\bfseries{Definition 2.}} If $u$ can only provide partial credential factors needed for one  arbitrary authentication path of $v$ with the attack profile, $u$ is then called the half capacity parent node of the node $v$.

{\bfseries{Definition 3.}} If $u$, $w$ or the other nodes can jointly provide all the credential factors needed for at least one authentication path of $v$ with the attack profile, $u$, $w$ and other nodes are then called the couples nodes to each other. The directed edges from $u$ to $v$ and from $w$ to $v$ are called weak-directivity edges.  

Based on the aforementioned definitions, We can formulate the correlation between the online accounts and the created edge can be divided into strong-directivity one and weak-directivity one. 
 To conduct the correlation formulation. briefly, we identify $NVP_{i}$, namely the number of authentication paths of each online account $oa_{i}$, we extract the credential factors needed for each path $cp_{in}$ in $oa_{i}$ and then match $cp_{in}$ with $pi_{j}$ from other $oa_{j}$. If $pi_{j}$ overlap $cp_{in}$ with attack profile (AP), then we created strong-directivity edge $se_{ji}$ between $oa_{j}$ and $oa_{i}$. We apply this operation to all the nodes in $OA$ to obtain all the strong-directivity edges and full capacity parents for node $oa_{i}$. Then we match the individual credential factors of $cf_{ni}$ to the non-full-capacity parent nodes,e.g., $oa_{c}$ and $oa_{d}$, and if $oa_{c}$ and $oa_{d}$ can provide personal information overlap $cp_{ni}$ with attack profile (AP), then we created weak-directivity edge $we_{cn}$ between $oa_{c}$ and $oa_{n}$ as well as $we_{dn}$ between $oa_{d}$ and $oa_{n}$. We define the Couple File($CouF$) to record the tuple $(oa_{c},oa_{d};oa_{n})$ which refers that $oa_{c}$, $oa_{d}$ are couple nodes to $oa_{n}$. Finally, we obtain the strong-directivity and weak-directivity among all the online accounts in the Transformation Dependency Graph.


\subsection{Strategy Output}

 On the basis of the Transformation Dependency Graph and correlation of online accounts, we develop the strategy engine to  explore the potential attack paths of the online account ecosystem. The strategy engine provides online account providers with querying the attacks in two scenarios.  The first one is when given a set of personal information as credentials in addition to SMS Codes (or the accounts that the attacker has compromised), the system should be able to output the possible accounts that can be attacked within the Online Account Ecosystem. This may occur when the attacker obtains the cellphone number and SMS Code or when the data breach happens in the Internet initially. The other scenario is when an attacker aims at a specific target account with its required reset credentials, the possible attack path starting from low-security-level accounts needs to be provided.


To solve the first question, we define the initial attacked online accounts as Online Account Attacked Set ($\mathit{OAAS}$). Then we collect all of the personal information of $\mathit{OAAS}$ as an Initial Attack Database ($\mathit{IAD}$). Then we compare credential factors of each online account with factors in the $\mathit{IAD}$. If all credential factors of one authentication path can be found in the $\mathit{IAD}$, we consider that this online account can be attacked on this condition. Then we add the personal information from the attacked online account to the $\mathit{IAD}$ and repeat the previous processes. Finally, we get all the potential account victims ($\mathit{PAV}$).

To solve the second question of attacking the specific online account ($\mathit{oa_{t}}$), we assume that attacker owns the ability to obtain cellphone number and intercept SMS Code. We initially search the full capacity parent nodes and the half capacity couple nodes of the target accounts. In addition, we merge each group of half capacity couple nodes of the target account as one full capacity node. If the credential factors of the full capacity parent nodes and the merged couple nodes are not cellphone number plus SMS Code, we iteratively assign the full capacity parent nodes and the merged couple nodes as target account and repeat the aforementioned process. When we meet the nodes whose credential factors are cellphone number plus SMS Code, We stop searching and return the account chain. 

%% file: Measurement.tex
\section{Measurement Study}
\subsection{Measurement Setup}

To evaluate the performance of ActFort, we conducted extensive experiments on 201 representative Internet services (top-ranked on www.alexa.com) both for the websites and for their mobile applications (if any).

The key aspects we investigated are summarized as follows:
\squishlist
	\item What are the credential factors used for authentication and their proportion in Online Account Ecosystem? 
	\item What are the types of personal information that can be obtained from the online accounts and their proportion in Online Account Ecosystem? 
	\item What are the most vulnerable online accounts within Online Account Ecosystem? 
\squishend

We manually set up test accounts and collected all possible Authentication Process methods and types of personal information leaked for all the services we investigated. Using Transformation Depency Graph Generation, we automatically generated dependency graphs and obtained analysis and strategies. All the proposed attack paths were verified by us later. 
We also spilt our measurement into different domains (e.g. Fintech, Email, Social Network, etc.) to have a better understanding of different sections in Online Account Ecosystem.  

\subsection{Evaluation}
\subsubsection{Measurement Results}
\begin{figure}[h]
	\centering
	\includegraphics[width=0.46\textwidth]{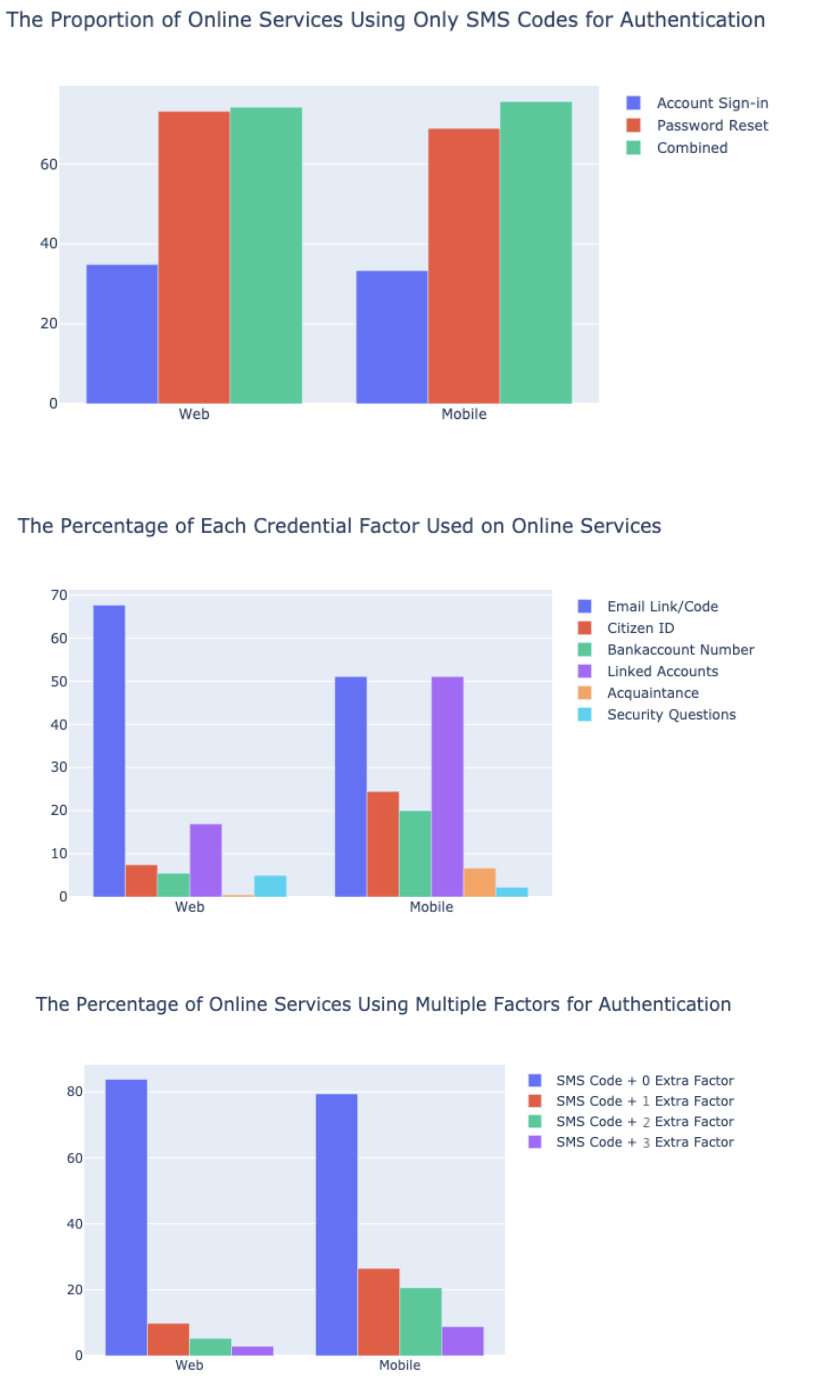}
	\setlength{\abovecaptionskip}{-1pt}
	\caption{The Measurement Results for Authentication Process} 
	\label{Fig:result_bar}	
\end{figure}

 According to our analysis, there are 405 authentication paths in total. Figure~\ref{Fig:result_bar} shows the proportion of online accounts whose authentication only involves SMS Code. Figure~\ref{Fig:result_bar} also lists the proportion of other common credential factors used for online accounts. First, we note that the percentage of services using merely SMS codes for sign-in is significantly lower than for password resetting, which implies that attacking accounts using password resetting is easier.  Additionally, according to our study, SMS Code takes up over 80\% for the authentication and less than 20\% website services demand extra information to authenticate users. 

Furthermore, we found that these authentication paths can usually be divided into 3 types: general path (45\% on the mobile applications while 58.65\% on the websites), which uses basic authentication factors; info path (17\% on the mobile applications while 13.45\% on the websites), which requires factors like real names and phone numbers; unique path (17\% and 16.35\% on the mobile applications and the websites respectively), which uses factors like biometrics.

We also analyzed the personal information in the user interface in each online account after login. We found that varieties of information exposed on websites interface were similar to that on the mobile application. Table~\ref{tab:e4} shows that the top three of high percentile personal information are cellphone number, user real name, and email address. Most of the personal information appeared exceeds 40\% except the device type and the bankcard number. This indicates that these online services pay insufficient attention to protecting users' most privacy information on the user interface. 
However, unsurprisingly, protection of bankcard number is better than any of others. Although none of the online accounts expose the whole binding bankcard number, masked digits of bankcard number are inconsistent in different online accounts, which will be discussed later. 

\begin{table}[h]
	\centering
	\setlength{\abovecaptionskip}{-1pt}
	\caption{The percentage of Private Information Obtained From Online Accounts After Log-in.}
	\label{tab:e4}
	\footnotesize 
	\begin{threeparttable}
		\begin{tabular}{|c|c|c|c}
			\hline			
			\makecell[c]{\bfseries Credential Factors }&\makecell[c]{ \bfseries Web Account.  /\%}&\makecell[c]{ \bfseries Mobile Account.  /\%}\\
			\hline 
			\bfseries Real Name & \makecell[c]{49.20} & \makecell[c]{ 75.00} \\
			\hline 
			\bfseries Citizen ID & \makecell[c]{11.76} & \makecell[c]{ 41.07}  \\
			\hline 			
			\bfseries Cellphone Number & \makecell[c]{54.01} & \makecell[c]{87.50}  \\
			\hline
			\bfseries E-mail Address & \makecell[c]{59.36} & \makecell[c]{64.29}  \\
			\hline
			\bfseries Address & \makecell[c]{51.34} & \makecell[c]{64.29}  \\
			\hline
			\bfseries User ID & \makecell[c]{45.99} & \makecell[c]{60.71}  \\
			\hline
			\bfseries Binding Account & \makecell[c]{44.92} & \makecell[c]{57.14}  \\
			\hline
			\bfseries Acquaintance Info. & \makecell[c]{32.09} & \makecell[c]{66.07}  \\
			\hline
			\bfseries Device Type & \makecell[c]{14.97} & \makecell[c]{35.71}  \\
			\hline
		\end{tabular}
	\end{threeparttable}
\end{table}

After obtaining credential factors and personal information from over 200 online services. We generated the dependency graph and explore the correlation among these online accounts. We found that there are four types of dependency relationship among all the evaluated accounts: (1) the account can be directly compromised with attack profile (cellphone number and SMS Code), (2) the account can be compromised with one layer middle account, (3) the account can be compromised with two-layer middle accounts (all full capacity parents), and (4) the account can be compromised with two-layer middle accounts (full capacity parents and half capacity parents.
According to the evaluation on the online services, 74.13\% of the website and 75.56\% of the mobile applications can be compromised with the cellphone number and SMS Codes directly. 9.83\% and 26.47\% of the accounts can be compromised with one-middle-layer accounts respectively for websites and mobile apps. 5.20\% of the website accounts and 20.59\% of the mobile application accounts can be compromised with with two-layer middle accounts (all full capacity parents). Percentages of  accounts can be compromised with with two-layer middle accounts (full capacity parents and half capacity parents) are 2.89\% and 8.82\% . Only 4.44\% for website accounts and 2.22\% for mobile application accounts can't be compromised with the evaluated accounts. Note that the overall percentage can not be summed up to 100\% since one service can have multiple reset combinations.

\begin{figure}[h]
	\centering
	\includegraphics[width=0.4\textwidth]{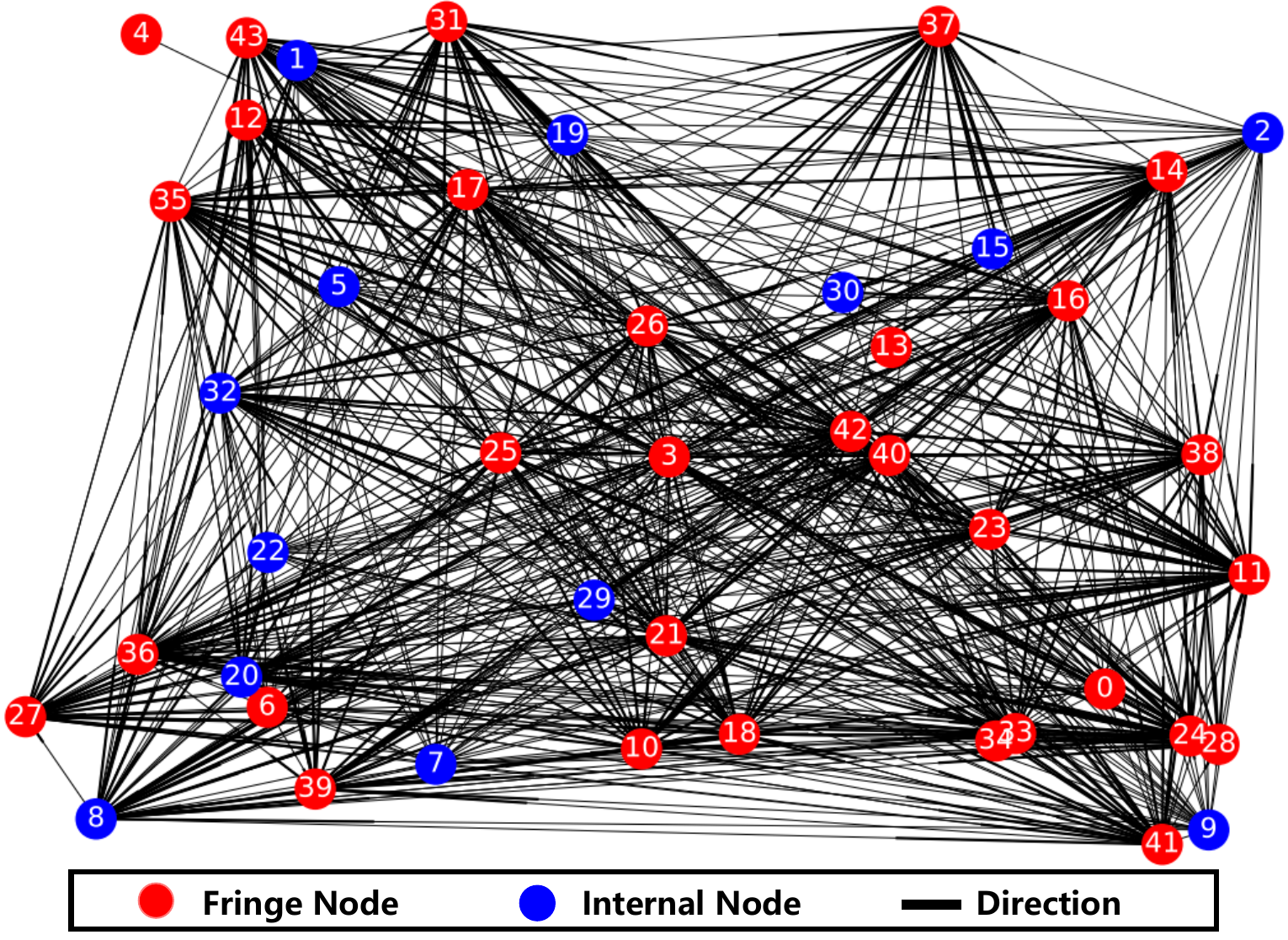}
	\setlength{\abovecaptionskip}{-1pt}
	\caption{Connections Among 44 Online Accounts}
	\label{Fig:nodes}	
\end{figure}

Figure~\ref{Fig:nodes} shows the connection graph of 44 online accounts. Red dots represent the fringe nodes, i.e., the online accounts which only need cellphone plus SMS Code for authentication. Blue dots represent the internal nodes, i.e., the online accounts which  need extra credential factors except SMS Code. The directed edge represents that the source nodes can provides all the credential factors for the destination nodes besides cellphone number plus SMS Code.  

On the basis of the generated graphs, we further explore the vulnerability of the nodes. As mentioned above, we could intercept SMS Codes sending to one specific phone such that we could login these accounts can be verified with only SMS Code and phone number. According to collected data, we found that authentication with email code/link was the most common method except with SMS Code. If we want to attack internal nodes in the graph as many as possible, at first, we ought to attack email account. We investigated the most common email accounts, i.e., Gmail.com, 163.com, Outlook.com, aliyun.com, and found that all of these accounts could be verified with only SMS Code. It means that mainstream emails have a low level of security so that attackers could verify victims' email accounts easily and they could verify a large proportion of accounts that belong to general-path mentioned previously. We further found that 3 websites, i.e., Ctrip, China Railway (12306) and Xiaozhu, gave the whole or vital part of citizen ID. Ctrip and Xiaozhu among these three could be verified with only SMS code or email code, so we could get citizen ID easily. In addition to this, we investigated accounts which showed device type of user and acquaintance names on their user interface and found that JD and Linkedin provided a mass of this type of information. Unsurprisingly, these accounts could also be verified with only SMS Code or email code. In the end, we investigated the authentication data of mainstream cloud storage service accounts, i.e., Baidu Pan and Dropbox. We could use email code to verify Dropbox account and use SMS Code or email code to verify Baidu Pan account. On account of that many people were used to copy their photos as backups storing on cloud, we might get users' photos, users' citizen ID photos or other privacy as attack tools after attack cloud storage service accounts successfully. After these four steps, we could verify the majority of accounts. Later in our case study, we will demonstrate attacks on those core nodes and further finish the attack on Fintech services.



\subsubsection{Key Insights}
According to the results of the measurement, we discover some interesting insights that explain the current situation of Online Account Ecosystem. The corresponding countermeasures will be discussed in the next section.  
\squishlist
    \item{\bf{Emails are the gateway to the most of the vulnerabilities exposed.}} Most Email accounts can be reset merely using SMS Codes. It is common that users use Email addresses to sign up for other services and a compromised Email account can be a gateway for attackers to conduct further attacks. From the Email history, there is a high possibility that Email accounts will reveal important information, such as signed-up services, social relationship and recent activities, which can be used to set up a subtle attack.
    \item{\bf{Asymmetry exists between mobile Apps vs websites and sign-in vs resetting passwords.}} Several instances we found during the measurement suggest the asymmetry between mobile Apps and websites. For example, the difficulties of authentication processes on Alipay's web end and mobile end differ, where web end requires a more hard-to-get factor(namely bankcard number). Also, Gome (one of the largest electrical appliance retailers in China) covers part of users' SSN on web end while exposes the same part of the information on the mobile end. Similarly, there are different requirements for sign-in and resetting passwords, which should be equally secure as for attackers.
    \item{\bf{Different domains have different levels of authentication.}} Generally, Fintech services are deployed with the most strict authentications, which is also the main target for attackers to cause direct financial loss. To compromise a Fintech account, the attacker might need to collect as much personal information from other accounts as possible, sometimes even with the help of other illegal means like black market.
    \item{\bf{There is no unified rule for sensitive information/privacy protection in user info page.}} Different Internet service providers cover different parts of SSN and bankcard numbers on their accounts. By attacking several service accounts and applying certain combining rules, the attacker could easily cipher covered SSN and bankcard numbers.
    \item{\bf{Biometric features and U2F security key are the most secure authentication.}} A lot of highly-secure services involve U2F security key or biometric authentication using features like fingerprints and facial features. This is usually the most robust node in Online Account Ecosystem, which is hard for attackers to mimic.
\squishend

%% file: CaseStudy.tex
\section{Attack Studies}

In this section, we demonstrate how to conduct Chain Reaction Attack with the help of our system. To achieve our goal, Chain Reaction Attack involves 3 major steps, namely,attack path generation, SMS Code interception and high-value account intrusion. Note that we conducted our measurement throughout a long period of time and some of the cases shown here may not be consistent with the latest versions of some applications, but the key idea of Chain Reaction Attack can be demonstrated here and can be extended to other attack surfaces in the future for latest versions. 

\subsection{Attack Steps}
\subsubsection{Attack Path Generation}
The attacker can either choose random targets or conduct precise attacks on specific targets. Both attacks require victims' cellphone numbers and addresses(which means the attack would be conducted physically within a certain range of victims, see Section~\ref{threat}).
The difference between the two target selections lies in the way of obtaining the cellphone numbers of victims. When the attacker chooses the specific victim, he could utilize the existing illegal databases of leaked personal information to get the victims' information~\cite{express}. While in a random attack, attackers can deploy phishing Wifi at airports and railway stations to get surrounding potential victims' phone numbers.

\subsubsection{SMS Code Interception}
\begin{figure}[pt]
	\centering
	\includegraphics[width=6.8cm,height=5cm]{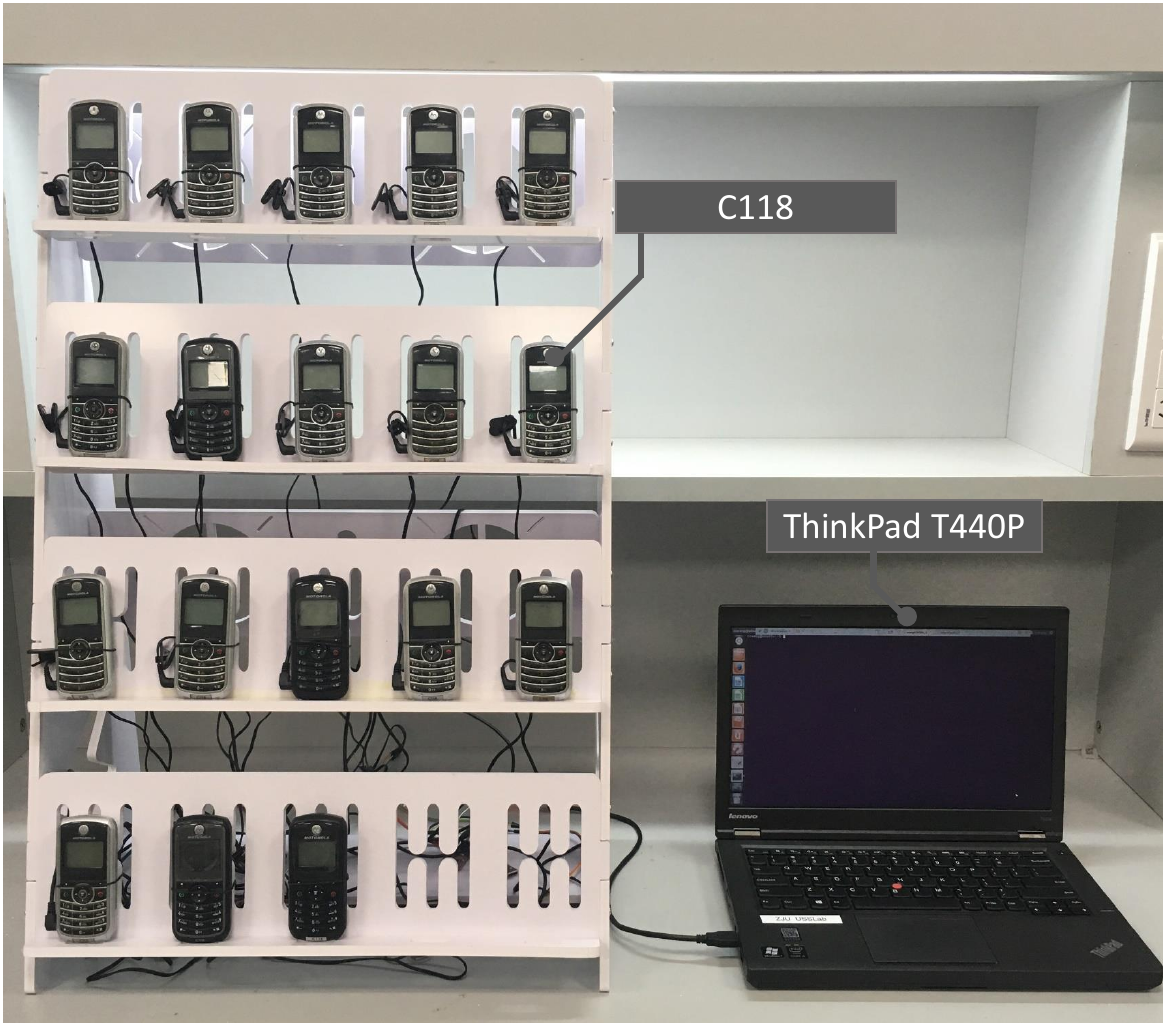}

	\setlength{\abovecaptionskip}{-1pt}
	\caption{Experimental setup for SMS sniffing attack.} 
	\vspace{-0.5cm}
	\label{Fig:device1}	
\end{figure}

\begin{figure}[pt]
	\centering
	\includegraphics[width=0.42\textwidth]{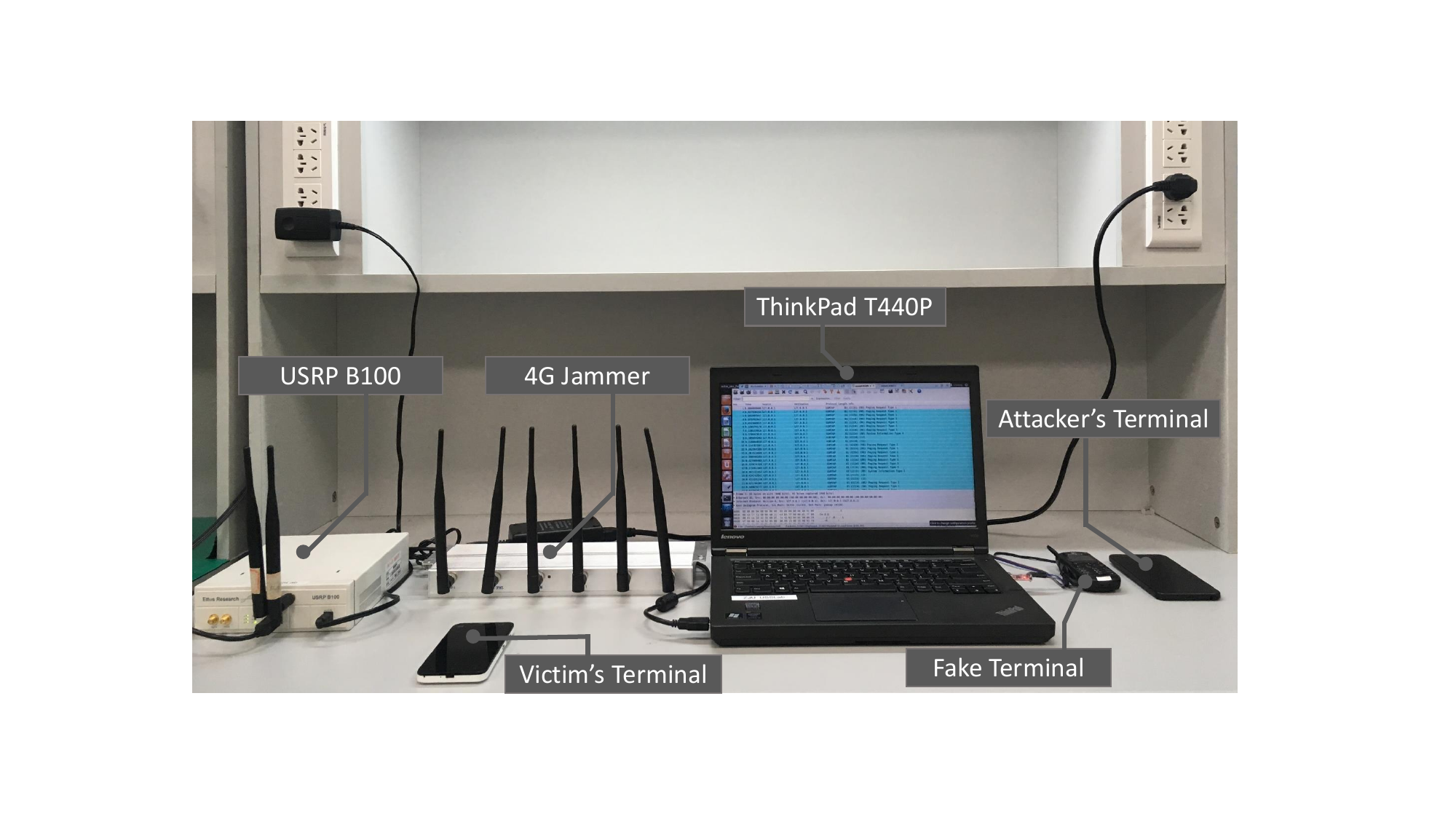}

	\setlength{\abovecaptionskip}{-1pt}
	\caption{Experimental setup for the active MitM attack.} 
	\vspace{-0.5cm}
	\label{Fig:device2}	
\end{figure}

There are two feasible ways to intercept the SMS codes. The first one is low-cost Simple GSM Sniffing, as shown in Figure~\ref{Fig:device1}. In comparison, the second method, Active MitM Attack could strengthen the attacker's concealment but with a relatively high cost and complex equipments, as shown in Figure~\ref{Fig:device2}. Note that during our case study, we chose the simple and low-cost GSM Sniffing using OsmocomBB with cheap equipment Motolora C118 in the GSM network. Active MitM Attack can be realized using fake base stations powered by USRP ~\cite{dubey2016demonstration} after the LTE network is downgraded to GSM forced by a 4G jammer ~\cite{huanglte}. 

 OsmocomBB~\cite{osmocom} is a free open source GSM baseband software implementation and it intended to completely replace the need for a propriety GSM baseband software. However, attackers can leverage it to intercept the SMS messages with specific firmware (e.g. Motolora C118).
Although GSM has its own encryption algorithm for SMS transmission, many GSM networks have no or weak data encryption. If the SMS transmission is encrypted with A5/1 algorithm~\cite{biham2000cryptanalysis}, existing hacking method~\cite{a51} can be used to obtain the session key.

We used one Thinkpad T440p model laptops, with Intel i5 4200M CPU and 4G RAM in the experiments. We connected 16 customized C118 cellphones to the laptop over USB cables. Each C118 could monitor one frequency point in the GSM network. We ran the customized OsmocomBB software to decode the GSM signals and Wireshark to filter the target SMS Codes with specific rules. Once the steps above were successfully completed, we could sniff SMS Codes transmitting through several nearby base stations. One caveat is that the victim can also receive the SMS Code and mitigate the stealthiness of the attack. Picking a perfect timing when victims' vigilance is at a relatively low state like midnight could solve this problem in some degree.



\begin{figure}[!t]
	\begin{center}
    \includegraphics[width=0.48\textwidth]{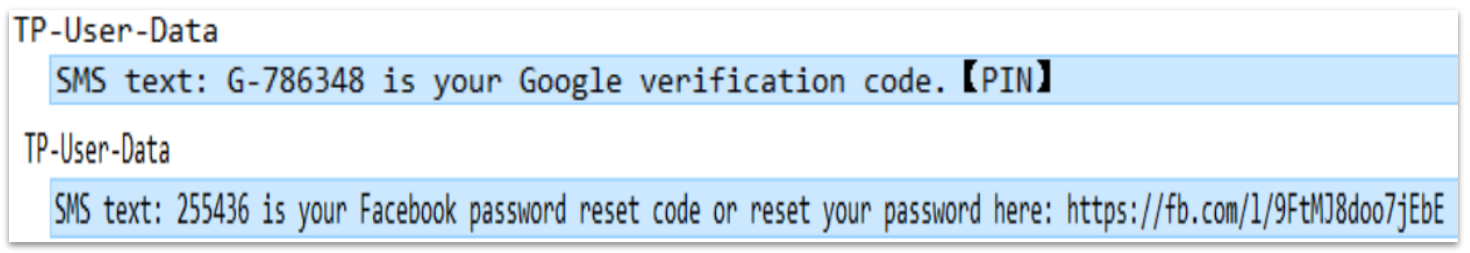}
	\end{center}
	\setlength{\abovecaptionskip}{-1pt}
	\caption{Example Intercepted SMS Code from Google and Facebook shown in the wireshark.}
	\label{Fig:sms}%
	\vspace{-0.5cm}
\end{figure}



\subsubsection{High-value Account Intrusion}
Many highly secure online account authentications and password resetting process request users to input not only SMS Code but also other credential factors. Therefore, to conduct the attack, the attacker must obtain extra personal information for authentication. The transformation dependency relationship of online accounts provides the attacker with possible attack paths. The attacker first compromises the least secure account(s). Subsequently, the acquired personal information from that account(s) helps facilitate an attack on another account. The follow-up accounts would disclose more vital information and enable further attacks.

\subsection{Real Cases}
 As mentioned previously, cellphone numbers and victim addresses could be easily obtained through a few illegal existing methods. We can also use Simple SMS Sniffing to intercept SMS codes. The key part here is to find the feasible attack paths using ActFort. The following real cases use the attack paths provided by ActFort. 

\noindent\textbf{Case I.} We used SMS code as a one-time token to directly log into Baidu Wallet account. Once we logged into the account, we were eligible to use QR code to make a payment in the app. There is no intermediate attack needed. 


\noindent\textbf{Case II.} We tried to attack Paypal accounts. To reset Paypal login password, both SMS code and Email code were required for resetting the password. So we chose Gmail as an example in this case. Users are allowed to use a phone number to login or reset password in Google login session. Using SMS code to reset the Gmail account linked to the phone number, we successfully logged into the Gmail account and got the email address with a temporary token sent by Paypal. Now we could use compromised Paypal account to make transactions. 

\noindent\textbf{Case III.} We deployed attacks on several Internet service providers to complete a chain attack. Alipay is one of the two mainstream mobile payment platforms in China. Note that unconformity of authentication factor combinations exists between  the mobile application and the web client of Alipay, so we conducted our attacks on both platforms separately.

On the mobile application, Alipay provided several SMS-based combinations to reset password, such as face scanning, bankcard information and related questions. Despite how thoroughly-covered and secure-looking those choices were, once the very vulnerability within them was spotted(for us it was using citizen ID in addition to SMS code), the whole defense collapsed. Similarly, we used Citizen ID and SMS Code to reset the payment code. Citizen ID information is severely leaked and commonly traded in the black market in China, but instead of getting the ID in this specific way, we attacked another website to obtain it. We attacked Ctrip, a leading online travel agency in China SMS code as a one-time login token. On the account page, we found user's citizen ID information by simply clicking "EDIT" button for Frequent Travelers Info. With the citizen ID and a SMS Code), we were able to reset Alipay's password and payment code and make transactions.
 
On the web client, there is an additional customer service option for resetting password. With the personal information gathered from other online services, this increases the attacker's chance of compromising the account by applying social engineering. 


%% file: RelatedWork.tex
\section{Related Work}

A few existing methods could be utilized to conduct attacks on a single account. Gelernter et al.~\cite{gelernter2017password} presented PRMitM, a way to reset victims' account using a malicious fake website and then collecting resetting challenges from victims.  Collin Mulliner et al.~\cite{mulliner2013sms} offered a trojan-based SMS attack on smartphones OS. By installing an SMS trojan on a victim's cellphone, every time the victim receives an SMS-based one-time password, the attacker is able to get it and use it to launch an account attack. Ghost Telephonist~\cite{zheng2017ghost} is an attack based on one vulnerability (the authentication step is missing) found in Circuit Switched Fallback in LTE networks, which enables the attacker to impersonate both the caller and the callee and could be considered as a tool for further attacks. With a large SMS dataset, Bradley Reaves et al.~\cite{reaves2016sending} analyzed the security of SMS as a verification method and unveiled the potentially sensitive information leakage when those popular companies built their services on the basis of SMS.

There are also attack studies on other factors like account linking features and social authentication (acquaintance) described in our measurement. Ghasemisharif et al. ~\cite{ghasemisharif2018single} investigated the flaws caused by Single Sign-On and explored potential attacks among binding accounts. This finding agrees with the result from our measurement that linked accounts could be disastrous. To break social authentication provided by companies like Facebook, Polakis et al. ~\cite{polakis2012all} implemented face recognition with cloud service to obtain victims' acquaintance information as sign-in factors.
In 2019, Doerfler et al. ~\cite{doerfler2019evaluating} conducted a measurement on the current online login challenges . Their study focused only on a single web service for the analysis. Although their results suggest that using factors like device authentication and SMS codes or even a combination of them are relatively secure against malicious account takeover, our Chain Reaction Attack can still compromise those secure accounts at a very high chance.

However, different from our Chain Reaction Attack based on ActFort, the attacks introduced above merely focus on a single account, which means that every attack on a different website is considered independent. Though, these attacks can be integrated in the intermediate steps in our Chain Reaction Attack to overcome the difficulty of obtaining some unique factors and make our attack more feasible under some special situations, but in comparison, our work exploits an interactive vulnerability at a system level and could be more explosive if used by real-world attackers.

%% file: Discussion.tex
\section{Discussion}
\label{sec:discussion}

\subsection{Countermeasures}
\subsubsection{Sensitive Information Protection Principles}
We proposed Online Account Information Protection Mechanism against Chain Reaction Attack according to the results of the measurement and the insights that we have. The key idea is to effectively cut some specific links in Information Flow with the help of our ActFort system. 

Considering the online account reciprocal transformation loophole pattern that we have obtained in the last section, we conclude our protection principals for online account information as follows: 
\squishlist
	\item {\bf{Cover unified digits on SSN and bankcard numbers.}} We propose that all the Internet service providers should cover their users' sensitive information and also cover them under a unified standard. By standardizing user information cover rules, the vulnerability of account interconnections within the Online Account Ecosystem will be alleviated.
	\item {\bf{Make email service accounts more secure.}} Email accounts serve as one of our breakthroughs during our attack. However, most email service providers, e.g., Google and NetEase (163), do not pay enough attention to their account security and can be attacked by simply resetting password via SMS codes. In order to enhance the security of the whole ecosystem, we strongly recommend that email service providers should bring their authentication method to a higher level.
	\item {\bf{Tackle the asymmetry existing between web end and mobile end.}} This kind of asymmetry should be avoided by developers. We reckon that within an enterprise there are usually different dev teams in the charge of different user terminals. Miscommunication can lead to asymmetry in security design, which expose user accounts to vulnerabilties. 
\squishend

\subsubsection{Built-in Authentication}

Due to the fragility of sending authentication codes via GSM SMS, it is necessary to find an alternative way to secure authentication code sending process. Companies like Duo~{\cite{duo}} and Tencent~{\cite{qq}} have designed applications to protect users' access to other applications and services.  There are also some services that start using instant message services like Whatsapp to deliver authentication codes. The user firstly is required to install a 3rd-party authentication application (3PAA). After installation and registration for 3PAA, the user needs to switch to the applications or services that are to be protected, finishing steps of authorizing 3PAA. Since 3PAA uses more secure protocols to transfer data, it is hard for attackers to decipher authentication codes. As long as user's 3PAA is not hacked, security of authentication via this way stands out. However, the trust towards the 3PAA from users and the inconvenience should be considered in the actual implementation.


\begin{figure}
	\centering
	\includegraphics[width=0.4\textwidth]{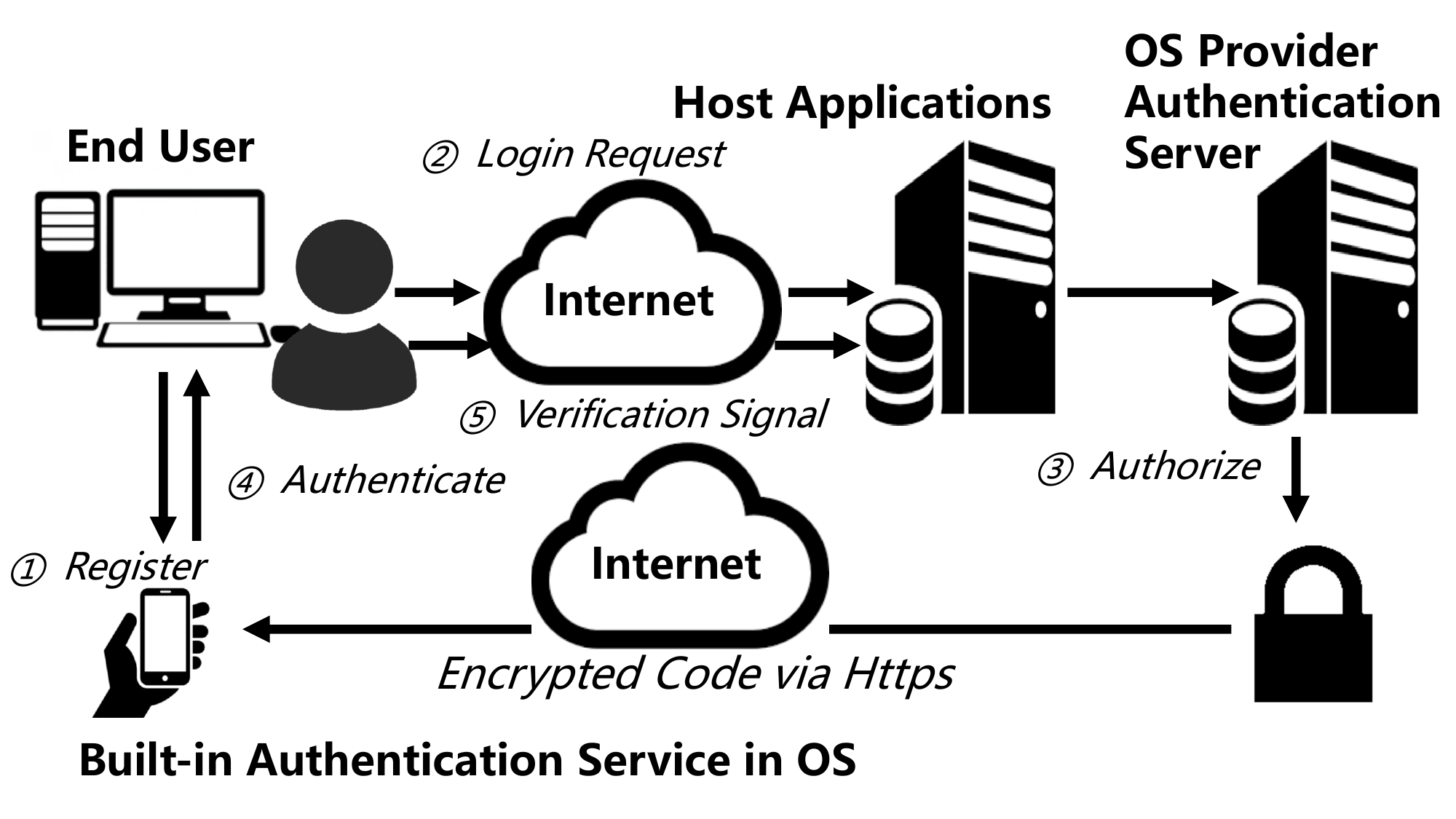}
	\setlength{\abovecaptionskip}{-1pt}
	\caption{Diagram of Built-in  Authentication Service in OS.} 

	\label{Fig:buildin}	
\end{figure} 

Hence, we propose a general diagram for Post-GSM built-in mobile authentication service in Fig.~\ref{Fig:buildin}. This is similar to what Apple has been doing. Apple has an option for users to get an authentication push on iOS devices for identifying themselves, which contains authentication attempt's location information. This function is embedded in iOS, which does not use GSM service. To go a step further to prevent malware code sniffing attack in mobile OS, we believe that there could be an industry standard that requests mobile device manufacturers or mobile telecom carriers to build an authentication push mechanism via encrypted dataflow into their devices or systems without actually displaying and saving codes in places like message inbox. Additionally, there could be a system level API for APP developers to directly call the authentication without user intervene which also improve usability. Users are more likely to trust services provided by the device manufacturers, the mobile telecom carriers and the original systems. 


\subsection{Limitations \& Future Work}
Chain Reaction Attack is a SMS-based attack from a view of the entire online account ecosystem. This attack possibly affects the majority of the current online services. However, it has several limitations.

Due to the current deployment of online authentications, our attack relies heavily on SMS code interception as its primary step in most attack cases. SMS sniffing first restricts the attack range to hundreds of meters of the victim. Phishing attack can solve this distance problem but also involves more social engineering effort and reduces the level of stealthiness.

Theoretically, any weak factors (like email code) in the ecosystem can be the breakthrough point, depending on the attacker's ability. This means Chain Reaction Attack is able to be expended to other forms of serial attacks beyond the our SMS-oriented attack.

Our ActFort measurement on top 200 web services does reveal several existing problems but hasn't covered a larger amount of accounts because some parts of our system still involve manual operations. Future work could further automatize the process to measure the ecosystem more thoroughly.

%% file: Conclusion.tex
\section{Conclusion}
\label{sec:conclusion}

In this paper, we introduce Chain Reaction Attack, which exploits vulnerabilities of interconnections and dependencies within Online Account Ecosystem. Taking SMS authentication code as the breakthrough of the attack, we are able to take full control of most online service accounts by resetting their passwords. To analyze the vulnerabilities systematically and provide website developers protection method for accounts, we design ActFort, a system that can automatically seek out dependencies among existing Internet service accounts. With the help of ActFort, we measure the mainstream Internet services and demonstrate real-world attack cases.

%% file: appendix.tex
\section{Appendix}

\subsection{SMS-based Authentication.}
\begin{figure}[h]
	\centering
	\includegraphics[width=0.35\textwidth]{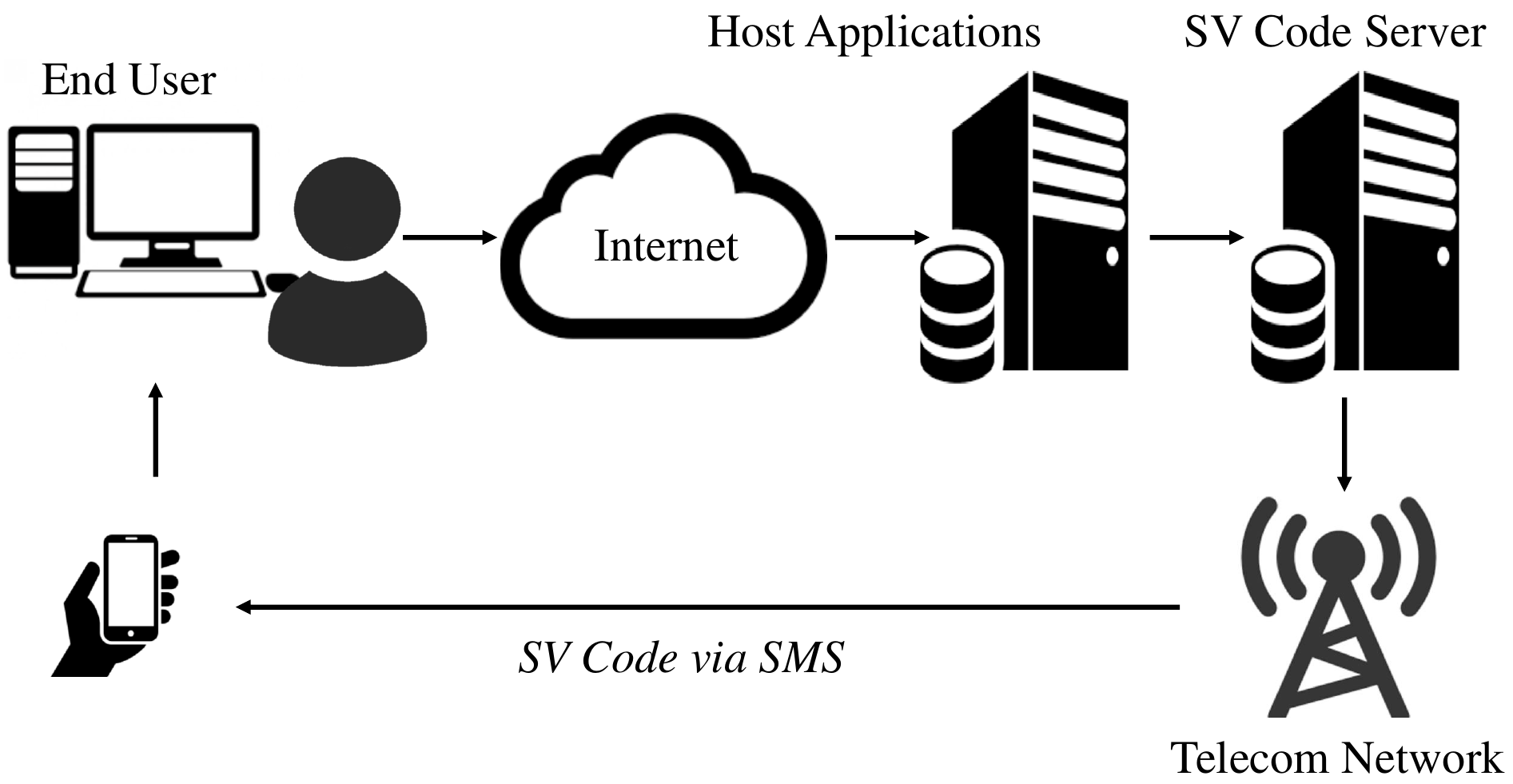}
    \setlength{\abovecaptionskip}{-1pt}
	\caption{Diagram of SMS-based Authentication.} 
	\vspace{-0.5cm}
	\label{Fig:smsauth}	
\end{figure}   

\subsection{Active MitM attack}
\begin{figure}[h]
	\centering
	\includegraphics[width=0.44\textwidth]{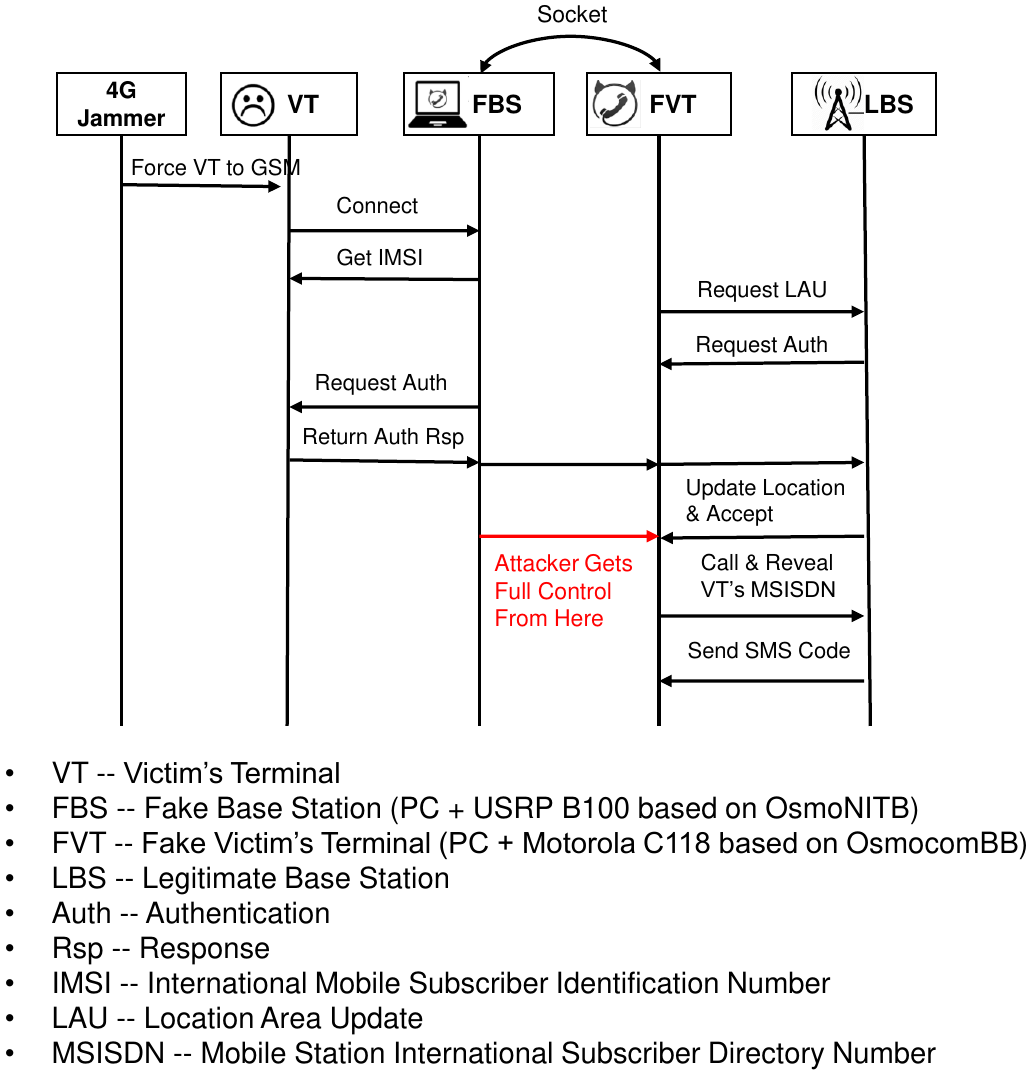}
	\caption{Active MitM attack} 
	\label{Fig:MitM_Appendix}	
\end{figure}

\subsection{Dependency Graph}
\begin{table}[h]
	\centering
	\setlength{\abovecaptionskip}{-1pt}
	\caption{Notations and definitions.}
	\label{tab:symbol}
	\footnotesize 
	\begin{threeparttable}
		\begin{tabular}{|c|c|}
			\hline	
			{\bfseries Symbol}&\makecell[c]{ \bfseries Definition } \\
			\hline 
			{$\mathit{OA=\left\{oa_{i}\right\}}$  }&\makecell[c]{Online accounts a victim owns.  }\\
			\hline 
			{$\mathit{CF=\left\{cf_{im}\right\}}$  }&\makecell[c]{Credential factors set---$m^{th}$ credential of $i^{th}$ account.   }\\
			\hline
			{$\mathit{PI=\left\{pi_{jn}\right\}}$  }&\makecell[c]{Personal Info. set---$n^{th}$ personal Info. of $j^{th}$ account.   }\\
			\hline
			{$\mathit{DE=\left\{de_{im}^{jn}\right\}}$  }&\makecell[c]{Directed edges set---Directed link from $cf_{im}$ to  $pi_{jn}$.   }\\
			\hline
			{$\mathit{VP=\left\{vp_{ik}\right\}}$  }&\makecell[c]{Verification paths set---$k^{th}$ path for $i^{th}$ account.   }\\
			\hline
			{$\mathit{CP=\left\{cp_{ik}\right\}}$  }&\makecell[c]{Credential factors needed for $\mathit{vp_{ik}}$.}\\
			\hline
			{$\mathit{SE=\left\{se_{ij}\right\}}$  }&\makecell[c]{Strong-directivity edges set.}\\
			\hline
			{$\mathit{WE=\left\{we_{ij}\right\}}$  }&\makecell[c]{Weak-directivity edge set.}\\
			\hline

		\end{tabular}
	\end{threeparttable}
\end{table}

\begin{figure}[h]
	\centering
	\includegraphics[width=0.44\textwidth]{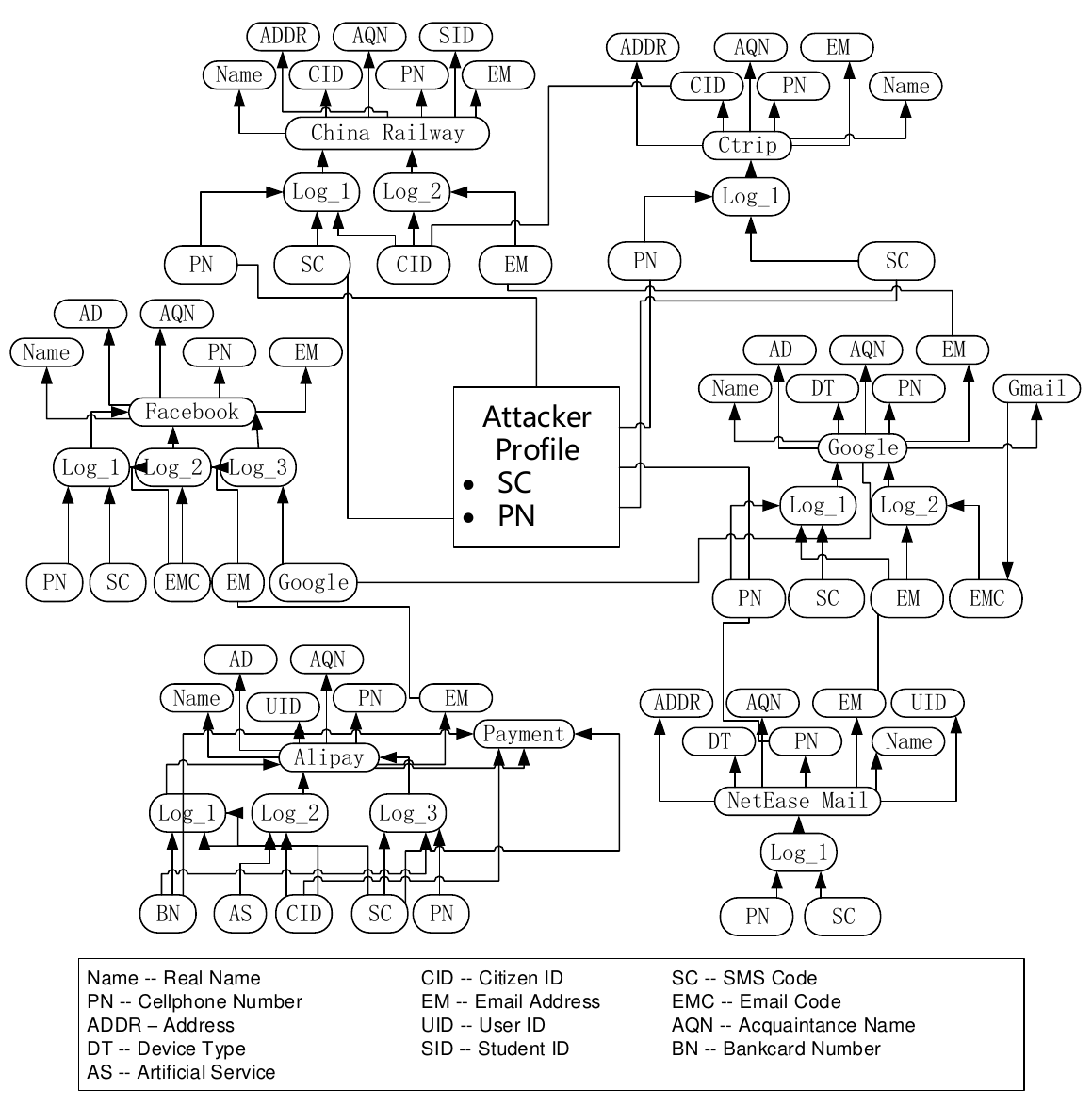}
	\setlength{\abovecaptionskip}{-1pt}
	\caption{Transformation Dependency Graph.} 
	\label{Fig:connection}	
\end{figure} 

\begin{figure}[h]
	\centering
	\includegraphics[width=0.35\textwidth]{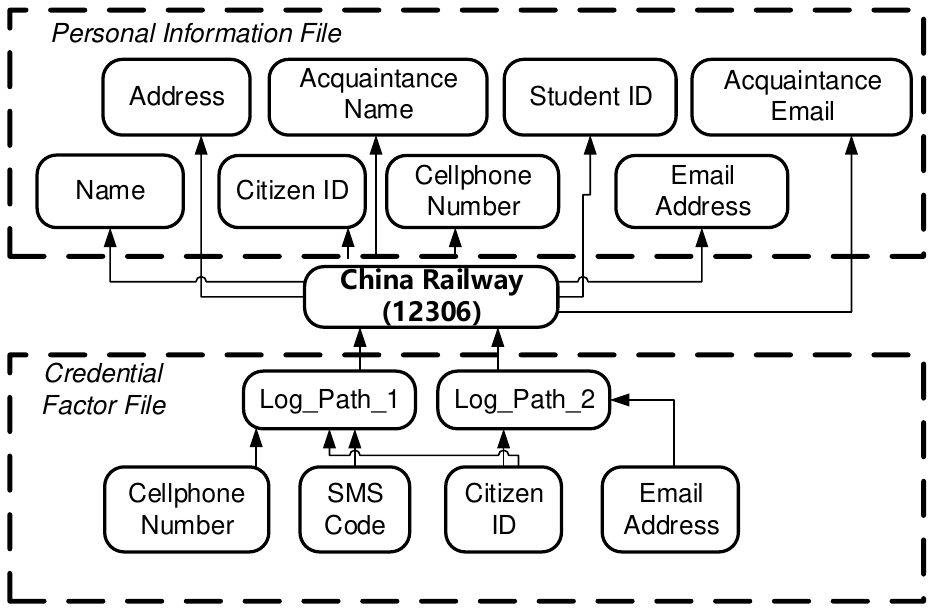}
	\setlength{\abovecaptionskip}{-1pt}
	\caption{Example Structure of One Single Node in TDG.}
	\label{Fig:DGG}	
\end{figure}

%% file: main.bbl
\begin{thebibliography}{10}
\providecommand{\url}[1]{#1}
\csname url@samestyle\endcsname
\providecommand{\newblock}{\relax}
\providecommand{\bibinfo}[2]{#2}
\providecommand{\BIBentrySTDinterwordspacing}{\spaceskip=0pt\relax}
\providecommand{\BIBentryALTinterwordstretchfactor}{4}
\providecommand{\BIBentryALTinterwordspacing}{\spaceskip=\fontdimen2\font plus
\BIBentryALTinterwordstretchfactor\fontdimen3\font minus
  \fontdimen4\font\relax}
\providecommand{\BIBforeignlanguage}[2]{{%
\expandafter\ifx\csname l@#1\endcsname\relax
\typeout{** WARNING: IEEEtran.bst: No hyphenation pattern has been}%
\typeout{** loaded for the language `#1'. Using the pattern for}%
\typeout{** the default language instead.}%
\else
\language=\csname l@#1\endcsname
\fi
#2}}
\providecommand{\BIBdecl}{\relax}
\BIBdecl

\bibitem{equi}
Equifax, ``Equifax,'' \url{https://www.equifaxsecurity2017.com/}, 2018.

\bibitem{Databreach}
csoonline, ``The 17 biggest data breaches of the 21st century,''
  \url{https://www.csoonline.com/article/2130877/data-breach/the-biggest-data-breaches-of-the-21st-century.html},
  2016.

\bibitem{owen2008multi}
W.~N. Owen and E.~Shoemaker, ``Multi-factor authentication system,'' May~13
  2008, uS Patent 7,373,515.

\bibitem{mulliner2013sms}
C.~Mulliner, R.~Borgaonkar, P.~Stewin, and J.-P. Seifert, ``Sms-based one-time
  passwords: attacks and defense,'' in \emph{International Conference on
  Detection of Intrusions and Malware, and Vulnerability Assessment}.\hskip 1em
  plus 0.5em minus 0.4em\relax Springer, 2013, pp. 150--159.

\bibitem{apvrille2010zeus}
A.~Apvrille, ``Zeus in the mobile (zitmo): Online banking’s two factor
  authentication defeated,'' 2010.

\bibitem{gelernter2017password}
N.~Gelernter, S.~Kalma, B.~Magnezi, and H.~Porcilan, ``The password reset mitm
  attack,'' in \emph{Security and Privacy (SP), 2017 IEEE Symposium on}.\hskip
  1em plus 0.5em minus 0.4em\relax IEEE, 2017, pp. 251--267.

\bibitem{barkan2005conditional}
E.~Barkan and E.~Biham, ``Conditional estimators: An effective attack on
  a5/1,'' in \emph{International Workshop on Selected Areas in
  Cryptography}.\hskip 1em plus 0.5em minus 0.4em\relax Springer, 2005, pp.
  1--19.

\bibitem{biryukov2000real}
A.~Biryukov, A.~Shamir, and D.~Wagner, ``Real time cryptanalysis of a5/1 on a
  pc,'' in \emph{International Workshop on Fast Software Encryption}.\hskip 1em
  plus 0.5em minus 0.4em\relax Springer, 2000, pp. 1--18.

\bibitem{golde2012weaponizing}
N.~Golde, K.~Redon, and R.~Borgaonkar, ``Weaponizing femtocells: The effect of
  rogue devices on mobile telecommunications.'' in \emph{NDSS}, 2012.

\bibitem{GuardianMobile}
guardian, ``Mobile web browsing overtakes desktop for the first time,''
  \url{https://www.theguardian.com/technology/2016/nov/02/mobile-web-browsing-desktop-smartphones-tablets},
  2016.

\bibitem{Wesoc}
wearesocial, ``Digital in 2018: World’s internet users pass the 4 billion
  mark,''
  \url{https://wearesocial.com/us/blog/2018/01/global-digital-report-2018},
  2018.

\bibitem{zviran1990comparison}
M.~Zviran and W.~J. Haga, ``A comparison of password techniques for multilevel
  authentication mechanisms,'' NAVAL POSTGRADUATE SCHOOL MONTEREY CA, Tech.
  Rep., 1990.

\bibitem{yan2004password}
J.~Yan, A.~Blackwell, R.~Anderson, and A.~Grant, ``Password memorability and
  security: Empirical results,'' \emph{IEEE Security \& privacy}, vol.~2,
  no.~5, pp. 25--31, 2004.

\bibitem{express}
I.~S. of~China, ``Investigation report on the protection of the rights of
  chinese netizens 2016,''
  \url{https://www.12321.cn/pdf/2016wangminquanyidiaochabaogao.pdf}, 2016.

\bibitem{dubey2016demonstration}
A.~Dubey, D.~Vohra, K.~Vachhani, and A.~Rao, ``Demonstration of vulnerabilities
  in gsm security with usrp b200 and open-source penetration tools,'' in
  \emph{2016 22nd Asia-Pacific Conference on Communications (APCC)}.\hskip 1em
  plus 0.5em minus 0.4em\relax IEEE, 2016, pp. 496--501.

\bibitem{huanglte}
L.~Huang, ``Lte redirection attack - forcing targeted lte cellphone into unsafe
  network,'' 05 2016.

\bibitem{osmocom}
Osmocom, ``Osmocombb project,'' \url{http://bb.osmocom.org/trac/}, 2013.

\bibitem{biham2000cryptanalysis}
E.~Biham and O.~Dunkelman, ``Cryptanalysis of the a5/1 gsm stream cipher,'' in
  \emph{International Conference on Cryptology in India}.\hskip 1em plus 0.5em
  minus 0.4em\relax Springer, 2000, pp. 43--51.

\bibitem{a51}
srlabs, ``A5/1 decryption,''
  \url{https://opensource.srlabs.de/projects/a51-decrypt/files}, 2010.

\bibitem{zheng2017ghost}
Y.~Zheng, L.~Huang, H.~Shan, J.~Li, Q.~Yang, and W.~Xu, ``Ghost telephonist
  impersonates you: Vulnerability in 4g lte cs fallback,'' in
  \emph{Communications and Network Security (CNS), 2017 IEEE Conference
  on}.\hskip 1em plus 0.5em minus 0.4em\relax IEEE, 2017, pp. 1--9.

\bibitem{reaves2016sending}
B.~Reaves, N.~Scaife, D.~Tian, L.~Blue, P.~Traynor, and K.~R. Butler, ``Sending
  out an sms: Characterizing the security of the sms ecosystem with public
  gateways,'' in \emph{Security and Privacy (SP), 2016 IEEE Symposium
  on}.\hskip 1em plus 0.5em minus 0.4em\relax IEEE, 2016, pp. 339--356.

\bibitem{ghasemisharif2018single}
M.~Ghasemisharif, A.~Ramesh, S.~Checkoway, C.~Kanich, and J.~Polakis, ``O
  single sign-off, where art thou? an empirical analysis of single sign-on
  account hijacking and session management on the web,'' in \emph{27th
  $\{$USENIX$\}$ Security Symposium ($\{$USENIX$\}$ Security 18)}, 2018, pp.
  1475--1492.

\bibitem{polakis2012all}
I.~Polakis, M.~Lancini, G.~Kontaxis, F.~Maggi, S.~Ioannidis, A.~D. Keromytis,
  and S.~Zanero, ``All your face are belong to us: breaking facebook's social
  authentication,'' in \emph{Proceedings of the 28th Annual Computer Security
  Applications Conference}, 2012, pp. 399--408.

\bibitem{doerfler2019evaluating}
P.~Doerfler, K.~Thomas, M.~Marincenko, J.~Ranieri, Y.~Jiang, A.~Moscicki, and
  D.~McCoy, ``Evaluating login challenges as adefense against account
  takeover,'' in \emph{The World Wide Web Conference}.\hskip 1em plus 0.5em
  minus 0.4em\relax ACM, 2019, pp. 372--382.

\bibitem{duo}
Duo, ``Duo,'' \url{https://duo.com/}, 2017.

\bibitem{qq}
Tencent, ``Tencent security,'' \url{https://aq.qq.com/cn2/index}, 2016.

\end{thebibliography}
